# Finding the Signal in the Noise: An Exploratory Study on Assessing the Effectiveness of AI and Accessibility Forums for Blind Users' Support Needs

Satwik Ram Kodandaram
Department of Computer Science
Stony Brook University
Stony Brook, New York, USA
skodandaram@cs.stonybrook.edu

Jiawei Zhou
Applied Math and Statistics
Stony Brook University
Stony Brook, New York, USA
jiawei.zhou.1@stonybrook.edu

Xiaojun Bi
Department of Computer Science
Stony Brook University
Stony Brook, New York, USA
xiaojun@cs.stonybrook.edu

IV Ramakrishnan
Computer Science
Stony Brook University
Stony Brook, New York, USA
ram@cs.stonybrook.edu

Vikas Ashok
Department of Computer Science
Old Dominion University
Norfolk, Virginia, USA
vganjigu@odu.edu

## Abstract

Accessibility forums and, more recently, generative AI tools have become vital resources for blind users seeking solutions to computer-interaction issues and learning about new assistive technologies, screen reader features, tutorials, and software updates. Understanding user experiences with these resources is essential for identifying and addressing persistent support gaps. Towards this, we interviewed 14 blind users who regularly engage with forums and GenAI tools. Findings revealed that forums often overwhelm users with multiple overlapping topics, redundant or irrelevant content, and fragmented responses that must be mentally pieced together, increasing cognitive load. GenAI tools, while offering more direct assistance, introduce new barriers by producing unreliable answers, including overly verbose or fragmented guidance, fabricated information, and contradictory suggestions that fail to follow prompts, thereby heightening verification demands. Based on these insights, we outlined design opportunities to improve the reliability of assistive resources, aiming to provide blind users with more trustworthy and cognitively-manageable support.

### ACM Reference Format:
Satwik Ram Kodandaram, Jiawei Zhou, Xiaojun Bi, IV Ramakrishnan, and Vikas Ashok. 2026. Finding the Signal in the Noise: An Exploratory Study on Assessing the Effectiveness of AI and Accessibility Forums for Blind Users' Support Needs. In *Proceedings of the 2026 CHI Conference on Human Factors in Computing Systems (CHI '26), April 13–17, 2026, Barcelona, Spain.* ACM, New York, NY, USA, 20 pages. https://doi.org/10.1145/3772318.3790956

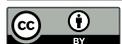

## 1 Introduction

Blind users routinely turn to online resources to solve their computer interaction problems, learn about new technologies that are relevant to them, and carry out routine digital tasks [21, 43, 52, 74, 105, 110, 114]. They draw on a broad ecosystem of mainstream materials such as official documentation (Microsoft Word [2]), blogs (Medium [1]), video tutorials (YouTube [5]), and Q&A platforms (Quora [3]), that offer guidance for troubleshooting and skill development. Yet, despite this apparent variety, these resources are often not accessible to blind screen reader users. Prior research has found that many technology tutorials are highly visual, depending on screenshots, spatial layout descriptions, and mouse-oriented "point-and-click" instructions that presuppose visual literacy and sighted usage patterns [59, 61]. For instance, the official Microsoft Word instructions for inserting a table[1] direct users to "click Insert" and "select Table" via annotated screenshots, but do not provide equivalent keyboard shortcuts or screen-reader-compatible steps. Even when blind users can access the text, the instructions seldom translate into non-visual, operationalizable steps. Thus, although online information seems plentiful, its real-world usefulness for blind people remains limited.

To navigate these barriers, blind users have long relied on accessibility focused[2] online community forums such as JAWS for Windows (JFW) [58], NVDA groups [81], AppleVis [82], and r/Blind [97]. Prior work shows that these spaces provide highly relevant, screen-reader-centric advice rooted in lived experience and tacit know-how [73, 75, 100, 107, 108]. They also foster emotional safety and a sense of community; over 60% of blind and low-vision individuals favor peer support instead of help from sighted people to avoid stigma, feelings of burden, or perceptions of inadequacy [12].

---

[1] https://support.microsoft.com/en-us/office/insert-a-table-a138f745-73ef-4879-b99a-2f3d38be612a

[2] Throughout the paper, we use "accessibility forums" and "accessibility-focused forums" interchangeably to refer to communities such as the JFW forum.

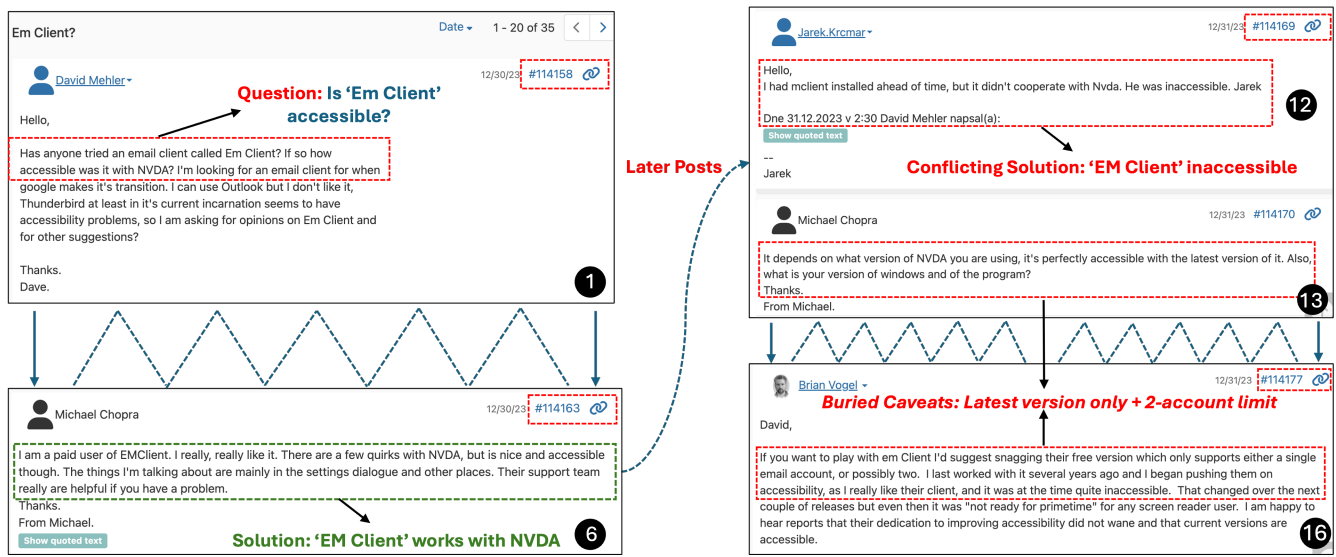

Figure 1: Example NVDA forum thread on "Em Client." The discussion begins with a potential solution, but crucial caveats and conflicting replies appear only after several intervening posts on unrelated topics. This scattered flow forces blind users to spend extra time piecing details together or risk overlooking them. Numbers (1, 6, 12, 13, 16) indicate the relative order of posts within the thread.

Recently, generative AI (GenAI) tools have introduced a new layer into this help-seeking ecosystem. Unlike traditional resources that require blind users to search across several inaccessible sites, GenAI systems consolidate information from several sources, including documentation, tutorials, blogs, and online examples, into a single conversational interface. These tools can summarize inaccessible content, reinterpret visually-oriented instructions, and translate them into sequential keyboard-based workflows, functioning as an on-demand support resource. For blind users, this consolidation reduces the friction of navigating scattered knowledge sources and provides rapid, personalized guidance.

Although prior research has extensively investigated the accessibility and usability of online information resources for blind and low-vision users, including websites, web applications, and online help communities [49, 51, 104], the ways in which blind users use accessibility-focused forums and GenAI tools to address their specific computer-interaction-related problems remain understudied. In particular, previous studies on accessibility forums have largely examined the structure and linguistic features of forum threads, documented the ways in which community members provide peer support related to application-specific tools and workflows, and proposed navigation-oriented interface interventions that restructure or disentangle threads to support more efficient screen-reader use [11, 60, 99, 106, 111]. Similarly, studies on GenAI use by blind individuals have mainly explored everyday or productivity contexts, such as visual interpretation, object recognition, content creation, coding assistance, live video help, and creative image generation [6, 34, 42, 91, 96, 101, 107]. Yet little is known about how these resources support blind users' computer interaction-related tasks, such as troubleshooting software, customizing screen readers, or adapting to interface changes introduced by updates. In particular, existing research has yet to fully characterize how blind screen-reader users consume responses from forums and GenAI tools, translate them into actionable steps, and apply those steps to resolve their computer-interaction challenges.

To address this knowledge gap, we conducted an IRB-approved interview study with 14 blind participants who regularly use forums and generative AI tools to obtain advice and solutions for their computer-interaction challenges and related needs. Our study is guided by the following three research questions:

- **RQ1:** What are blind users' experiences and challenges in accessibility forums when seeking computer-interaction support?
- **RQ2:** What are blind users' experiences and challenges with GenAI tools when seeking support for their computer-interaction issues?
- **RQ3:** What needs and preferences do blind users have regarding accessibility forums and GenAI tools when seeking support related to computer interaction?

To address these research questions, we conducted an IRB approved, semi-structured interview study with 14 blind participants who had lived experience using accessibility forums and GenAI tools for computer interaction challenges and related needs. The study uncovered a range of insights specific to these resources, extending beyond the well-documented accessibility barriers of web content navigation. Notably, in accessibility forums, participants reported that relevant information was often scattered across multiple posts in threads (Figure 1), making it mentally difficult to piece together important details for a complete resolution. In contrast, challenges with GenAI tools often stemmed from visually-anchored instructions (e.g., "click the green button"), overly verbose or fragmented responses, broken tutorial links, and fabricated information. Drawing on these findings, we outline design considerations aimed at improving the reliability, usability, and overall effectiveness of both accessibility forums and GenAI tools for blind users seeking computer-interaction support and staying informed about new assistive technologies.

In sum, this paper makes three key contributions:

(1) **Experiences and challenges.** Our study sheds light on the challenges blind users experience when relying on accessibility forums and GenAI tools to seek support for their computer-interaction needs.
(2) **Needs and preferences.** We document the specific needs and preferences of blind users when engaging with support platforms, including those related to step-by-step instructions, presentation of caveats, and overall trustworthiness of responses.
(3) **Design recommendations.** We propose actionable design suggestions for forums and AI systems, such as restructuring the threads, adopting accessible Q&A formats, and incorporating multi-agent verification, that provide concrete pathways toward enhancing the reliability and usability of these platforms as computer-interaction support resources for blind users.

## 2 Background & Related Work

Our study is broadly related to three bodies of prior work: web foraging by blind users, accessibility-focused online forums, and the use of generative AI for accessibility. We briefly review each area below.

### 2.1 Web Foraging by Blind People

The process of seeking online support and troubleshooting-related information is often understood through Information Foraging Theory (IFT) by Pirolli and Card [94], which describes strategies to gather "maximum information" with "minimal effort" [93]. Blind people who use screen readers often expend significantly higher effort when searching for and accessing information [30, 62, 67, 68, 110]. Prior research has consistently reported differences in information-seeking behavior and efficiency between blind and sighted users [61, 63, 83, 118, 119]. Studies indicate that, for blind users, searching for relevant content on the web is often challenging. They tend to formulate longer, more precise queries, use fewer reformulations, and rely less on search-support features such as auto-suggestions or spell-check [22, 41, 98]. As a result, they are more likely to remain on the first page of search results because processing long lists of results through a screen reader is significantly slower than visual scanning.

Efficiency outcomes further illustrate these disparities. Only about 13% of blind users go beyond the first page, compared to 43% of sighted users, and they typically visit fewer links per query (approximately 4 versus 13) [53]. Navigating through search results using screen readers also requires substantially more time for blind users, who spend an average of 106 seconds compared to roughly 36 seconds for sighted users [48]. Most blind users (80%) open only the top two results, which may limit the breadth of information considered, and only around 39% of blind users report they almost always find what they seek, compared to around 90% of sighted users [41, 50].

Beyond search strategies, several studies have documented challenges rooted in the linear, auditory nature of screen reader interaction. Sighted individuals can swiftly skim page layouts, visually scan code blocks, and use spatial hierarchies such as headings, lists, or formatting cues to locate relevant information efficiently [8, 19, 64]. In contrast, blind users must navigate content sequentially, which constrains their ability to skim and often results in information overload on visually dense, content-heavy pages [18, 95, 112, 117, 121]. These constraints compound the time and cognitive effort required, particularly in technical troubleshooting contexts where rapid parsing of structured or formatted information is essential. To address these challenges, various approaches have been proposed, including non-visual skimming interfaces that improve reading speed and search efficiency [8, 10], AI-driven summarization tools that enable non-linear exploration of page structures [117], and web accessibility standards such as WAI-ARIA or LLM-based HTML restructuring that enhance screen-reader compatibility [15, 124]. Nonetheless, complex layouts, missing semantic labels (e.g., headings, landmarks), dynamic content (AJAX, pop-ups, overlays), and advertisements continue to impose significant barriers [20, 23, 24, 66, 79, 100, 120]

From a theoretical standpoint, these limitations compel blind users to expend significantly more effort "per unit of information". However, little is known about how these challenges shape their knowledge-seeking behaviors and experiences with alternative support resources, such as accessibility forums and GenAI tools, including the specific difficulties they face on these platforms.

### 2.2 Community Forums for Accessibility Support

An internet forum [115] is an online platform to exchange messages on shared topics, ideal for asynchronous knowledge sharing, peer support, and problem solving [17, 56, 111]. Prior research has explored its structure, user engagement, and outcomes across education and technical domains [14, 84, 90, 103]. In educational contexts, forums can mimic classroom interactions and improve peer feedback [31], and students who actively ask and answer questions often achieve higher final grades [44]. This effect was particularly pronounced during the COVID-19 shift to remote learning, benefiting disadvantaged students whose GPAs improved through forum participation. In technical domains, Hellman et al. [54] found that users typically initiate threads while experts respond, with discussions focusing primarily on problem solving. Similarly, Ahmed et al. [9] highlighted forums' role in open-source projects for reporting defects, proposing new features, and providing user support, often via peer or "super-user" contributions.

While this body of work highlights the effectiveness of forums for learning, support, and problem-solving, it has largely focused on sighted users and general-purpose forums.

In contrast, blind people rely on accessibility-focused forums, such as the JFW (JAWS for Windows) group [58], NVDA group [81], AppleVis [82], and Reddit's r/Blind [97], to troubleshoot computer interaction-related issues, explore new assistive technologies, and learn about screen reader features, tutorials, and software updates. Although blind users do use general-purpose forums, prior studies show that they consistently prefer accessibility-focused forums because these communities offer tailored and experience-based assistance that general platforms rarely provide [72, 73, 75, 100, 107, 108]. For instance, blind and low-vision (BLV) programmers often turned to accessibility-focused forums rather than mainstream programming Q&A sites to seek help with inaccessible software tools and workplace barriers because these spaces provided concrete, lived-experience guidance for navigating inaccessible software and sociotechnical challenges in mixed-ability workplaces [89]. Albusays et al. [12] similarly reported that more than 60% of BLV participants preferred seeking help in accessibility-focused forums rather than from sighted colleagues, often to avoid the perception of being less capable. Collectively, prior research shows that accessibility-focused forums are critical and trusted spaces for problem-solving and community support among BLV users.

Given their importance to blind users, accessibility-focused forums have been analyzed in terms of their content and interaction patterns. For example, Saha et al. [99] examined six months of posts (180 threads) from a Logic Pro + VoiceOver mailing list, identifying topics, how blind users formulated queries, and how community members structured responses. They reported four thread types (*navigation/interaction, reviews, promotions, advocacy*) and found that blind users typically explained their issues, actions tried, and system configurations, while helpers provided standard terminology, step-by-step directions, added missing context, and shared example files and helpful links. Similarly, Johnson et al. [60] analyzed Program-L, a BLV programming mailing list, coding 173 messages from 20 newcomers (2016–2020). These messages captured how novices requested help and what support they received. They identified four novice profiles (*community, domain, programming, accessibility*) and showed that effective exchanges relied on disability self-disclosure, concrete assistance, and supportive norms for empathetic, contextual help. In a related study, Venkatraman et al. [111] surveyed accessibility forum posts and found they feature higher lexical density, more descriptive action verbs, and more first-person pronouns than general forums, making their dense yet structured language easier to follow with a screen reader.

Beyond content and interaction patterns, existing research has also examined navigation challenges in accessibility forums and proposed ways to make thread navigation more efficient for screen reader users. For instance, Sunkara et al. [106] showed that linear thread navigation and redundant content hinder efficiency. Anand et al. [11] proposed *TASER*, which disentangles overlapping sub-conversations to ease screen reader navigation. However, the scope of our study is far broader. We investigate how blind users navigate fragmented threads, track scattered details across posts, make sense of the information, translate it into a sequence of operationalizable steps, and apply those steps to resolve their own computer interaction-related problems.

In summary, prior work has provided qualitative accounts of accessibility-forum threads, examining how blind users formulate questions, how help is provided, and how conversational norms structure peer support within application-specific communities. These works have also characterized the linguistic features of accessibility-forum posts and proposed interface interventions that restructure or disentangle threads to improve screen-reader navigation. In contrast, our study examines how blind users forage for forum information when troubleshooting a specific computer-interaction problem, including navigating fragmented threads, tracking details scattered across posts, synthesizing advice from multiple replies, translating it into operationalizable steps, and applying those steps to resolve the issue.

### 2.3 GenAI for Accessibility

The advent of GenAI tools, such as *ChatGPT* [85], *Gemini* [47], and *Claude* [38], has significantly transformed the way information is accessed and used online [35]. These GenAI tools perform laborious web browsing tasks, saving users time by retrieving and summarizing results from vast internet sources. This efficiency has led users to rely extensively on them for information, from general factual questions to troubleshooting complex issues [7, 76]. Studies highlight their use in brainstorming ideas [29, 71], coding support [16, 37, 70, 88, 128], and efficient information retrieval [122].

These tools are particularly valuable for blind users, supporting scene interpretation, object recognition, label reading, and independent task completion [4, 6, 33, 45, 77]. Building on these application-specific systems, researchers have examined GenAI's broader role in accessibility, including diary studies of MLLM-based visual interpretation apps [46], methods for tracing provenance and detecting errors in image descriptions [39], and techniques for making text-to-image generation more accessible [57].

Extending beyond these works, several studies document a variety of GenAI applications for blind users, including everyday assistance, productivity and writing support, programming and coding help, live-video and real-world guidance, visual interpretation, and creative or image-generation tasks. [6, 34, 42, 91, 96, 101, 107]. Specifically, for productivity work, screen-reader users rely on GenAI for drafting and editing but must contend with verification overhead, verbosity, and limited workflow integration [91]. In programming, coding assistants boost throughput but introduce burdens such as suggestion overload and context switching [42]. For real-world assistance, live-video features in *ChatGPT* help with static scenes but struggle with spatial guidance, object handling, and continuous understanding in dynamic tasks [34]. In sum, while these applications show GenAI's growing value across many aspects of blind users' lives, they also introduce additional burdens, most notably the need to verify outputs, manage interaction breakdowns, and navigate new forms of friction.

Previous studies have also examined the daily use of GenAI among blind users. For example, Adnin et al. [6] conducted an interview study with 19 blind people to understand how they use and make sense of GenAI tools, particularly *ChatGPT* and *Be My AI*, in their everyday practices, such as recognizing uncertain objects, and information retrieval tasks related to the preparation of copywriting materials, emails, and creative writing (e.g., stories, poems, songs). The findings revealed how blind users navigate accessibility issues, inaccuracies, hallucinations, and idiosyncrasies associated with these tools. Similarly, Tang et al. [107] showed how blind users approached information access in the context of everyday uncertainty, adopting a mindset of skepticism and criticality towards both GenAI tools and the information provided.

Although prior work has examined blind users' engagement with GenAI in everyday and productivity scenarios, ranging from content creation and visual interpretation to coding support, live video assistance, and creative image generation, there is limited understanding of how these tools address computer-interaction-related needs, such as troubleshooting software, configuring screen readers, or adapting to interface changes. This paper addresses this gap by investigating how blind users leverage both GenAI and accessibility-focused forums to navigate these specialized interaction challenges, shedding light on practices, strategies, and emerging frictions.

| ID | Age | Sex | Age of Onset | Primary SR | SR Expertise | Familiar Forums | Familiar Gen AI Tools | Frequency Forums / AI |
|---|---|---|---|---|---|---|---|---|
| P1 | 37 | Female | 3 | JAWS | Intermediate | Reddit, JFW, AppleVis | ChatGPT | Bi-Weekly / Daily |
| P2 | 45 | Male | 18 | VoiceOver | Intermediate | AppleVis, JFW, BlindAds | ChatGPT, Gemini | Weekly / Daily |
| P3 | 61 | Male | 18 | JAWS | Intermediate | Reddit, NVDA, JFW | Gemini | Weekly / Daily |
| P4 | 43 | Male | 5 | VoiceOver | Intermediate | Reddit, AppleVis, NVDA, JFW | ChatGPT | Daily / Daily |
| P5 | 66 | Female | Congenital | JAWS | Intermediate | JFW | ChatGPT | Daily / Daily |
| P6 | 59 | Male | 28 | VoiceOver | Beginner | AppleVis | Gemini | Daily / Daily |
| P7 | 51 | Female | 19 | VoiceOver | Intermediate | AppleVis | ChatGPT | Weekly / Daily |
| P8 | 64 | Male | 6 | JAWS | Intermediate | JFW | ChatGPT | Bi-Weekly / Daily |
| P9 | 66 | Male | 5 | JAWS | Expert | JFW | ChatGPT | Bi-Weekly / Daily |
| P10 | 52 | Female | Congenital | JAWS | Intermediate | JFW | ChatGPT | Bi-Weekly / Daily |
| P11 | 58 | Female | 28 | JAWS | Expert | JFW, Reddit | ChatGPT | Weekly / Daily |
| P12 | 68 | Female | 19 | JAWS | Intermediate | JFW, AppleVis | ChatGPT | Daily / Daily |
| P13 | 33 | Female | 17 | JAWS | Intermediate | AppleVis, JFW, Reddit | ChatGPT, Gemini | Weekly / Daily |
| P14 | 73 | Male | Congenital | JAWS | Intermediate | JFW, AppleVis | ChatGPT | Daily / Daily |

Table 1: Demographics of blind participants in the interview study. All information was self-reported. "Bi-Weekly" indicates a frequency of once every two weeks.

## 3 Interview Study Method and Materials

To understand the barriers blind screen reader users face when seeking support for computer-interaction tasks, we conducted an IRB-approved interview study with individuals who routinely use accessibility forums and generative AI tools as support resources.

### 3.1 Participants

We recruited participants from an existing contact list of prior IRB-approved studies. We contacted the prospective participants who had previously consented to be recontacted for future research and had expressed an interest in studies related to accessibility and technology use. We also employed snowball sampling [80], asking participants to share the study with eligible peers. In accordance with our IRB protocol, the outreach was conducted using each individual's preferred communication method (email or phone).

Eligibility criteria required participants to: (i) self-identify as blind or severely visually impaired and rely on screen readers for computer use; (ii) regularly use screen readers for web navigation; (iii) have experience with accessibility-related forums (e.g.,*AppleVis*) and GenAI tools (e.g.,*ChatGPT*, *Gemini*); and (iv) communicate in English. We excluded individuals with mild visual impairment, those lacking experience with forums or GenAI tools, and those under 18. Screening interviews confirmed eligibility.

In total, 14 participants completed the study. Participants ranged in age from 33–73 (M = 55.4, SD = 12.2), with 7 male and 7 female. All participants identified as blind or severely visually impaired, regularly used screen readers, and reported experience with both forums and GenAI tools in practice. No participant reported any physical or auditory impairments that interfered with their ability to browse the web using a screen reader. Participant demographics are shown in Table 1.

### 3.2 Interview Protocol

We designed a semi-structured interview protocol that balances structure with flexibility, enabling both comparative insights across participants and rich in-depth accounts of individual experiences. The guide incorporated both closed and open-ended questions, allowing exploration of unexpected yet relevant topics as they emerged. The protocol was developed iteratively. The initial draft drew on prior literature on blind users' engagement with technology, discussion forums, and GenAI tools. To ensure relevance and practical grounding, we conducted a consultative interview with our University's Accessibility Officer, a blind domain expert. This session was audio-recorded and transcribed, after which we reviewed the transcripts and annotated salient responses with reflexive notes.

Guided by this preliminary analysis and consultation with the Accessibility Officer, we refined the interview questionnaire to improve clarity, phrasing, sensitivity, and coverage of key themes.

After the first five interviews, the guide was revised again. Participants' responses were analyzed through iterative coding [28], with emerging codes grouped into broader categories. These insights informed further refinements to the questionnaire and the addition of new probes. Iterative adaptation of the interview protocol is a well-established practice in qualitative HCI research [12, 24, 26, 107]. The final protocol comprised three thematic blocks, each containing core questions supplemented by optional probes to encourage elaboration when needed. The topics and sample questions under each block were as follows:

(1) **Experiences with Forums (RQ1).** *Can you describe your experiences using forums when trying to find solutions for a computer-related problem? What difficulties do you encounter when perusing discussions within forum threads? Can you share specific examples of scenarios when using a forum felt particularly frustrating or unhelpful? How do you usually determine that you've found all the information you need in a forum discussion?*
(2) **Challenges with GenAI Tools (RQ2).** *Can you tell me about your experiences using generative AI tools when you need help with assistive technology or computer applications? What challenges or difficulties do you face with the answers you receive from these tools? Can you share specific examples when AI responses were unhelpful or didn't work for your needs? How do you usually decide whether an AI-generated answer is reliable, useful, or correct?*
(3) **Needs and Preferences for Future Improvements (RQ3).** *Thinking about both forums and GenAI tools, what changes would make them easier to use and more helpful for you? What new features or improvements would you most like to see in these platforms to better support your needs?*

### 3.3 Procedure

The interviews were conducted in person at times that suited each participant. Prior to the interview, participants were briefed on the study procedures and provided written informed consent. Participants were also informed of their right to skip questions or withdraw at any stage without penalty. Each session began with participants introducing themselves, including their educational and professional background, visual impairment history, preferred screen readers, and participation in forums and GenAI tools. This was followed by a brief semi-structured interview (questions in Section 3.2). Although topics were predetermined, both topics and discussions were adapted to each participant's reported forums and tools. Participants were encouraged to illustrate responses by demonstrating interaction challenges on these platforms and to suggest improvements. With the consent of the participants, all interviews were recorded and subsequently transcribed. The lead researcher also maintained field notes documenting non-verbal cues (e.g., pauses, hesitations, tone) and contextual observations. Each interview lasted about 45-60 minutes, and all conversations were in English. Upon completing the study, all participants received a compensation of $80 *USD* for participating in the study.

### 3.4 Data Collection and Analysis

Each session was audio-recorded and transcribed using *Zoom* software with participants' consent. Transcription errors, including misheard words, typos, and homophone substitutions, were manually corrected. Names were replaced with pseudonyms (e.g., *P1, P2*), and sensitive identifiers were removed. The transcripts produced 462 single-spaced pages (max 50 lines/page), accompanied by two pages of field notes. All data were securely stored on encrypted, password-protected drives accessible only to the research team.

We analyzed the qualitative interview data using a hybrid reflexive thematic analysis [25, 27, 78, 113], combining deductive (top-down) coding guided by our research questions with inductive (bottom-up) coding emerging from the data. This approach aligned analysis with prior work [6, 107] while remaining open to unanticipated insights.

Before coding, we created a small set of a priori categories (deductive codes), such as *Forum Navigation*, *Understanding & Applying Forum Suggestions*, and *AI Response Quality*, drawn from our interview guide and literature. Prior studies on discussion forums [11, 106] and GenAI tools like *ChatGPT* [6, 107] informed these scaffolding categories, which were not fixed themes but helped structure the analysis around our RQs.

The first author then reviewed all transcripts line by line, open-coding noteworthy responses and assigning both deductive and inductive codes. Deductive and inductive codes were merged into a unified codebook, which the co-authors reviewed alongside sample excerpts and refined through weekly discussions. When new codes were introduced, the first author revisited earlier transcripts to ensure consistency. All authors then met to discuss and collaboratively group related codes into higher-order themes, drawing on their perspectives and expertise in accessibility research to analyze relationships and derive broader insights.

## 4 Key Findings

We present the key findings of our interview analysis, illustrated with the verbal comments of the participants. As shown in Figure 2, the section is organized around major themes and sub-themes that emerged: challenges with accessibility forums, challenges with GenAI-based tools (e.g., *ChatGPT*), and blind users' needs and preferences, each directly addressing our research questions. Quotations are labeled with participant codes (e.g., *P1, P2*), and italicized text indicates direct quotes.

### 4.1 RQ1: Experiences and Challenges with Accessibility Forums

Participants consistently indicated that forums were their first stop for resolving computer issues; however, they frequently faced obstacles that made it difficult to find, understand, and implement solutions. These obstacles typically revolved around the recurring challenges outlined below.

*4.1.1 Difficulties Searching for Information Within Threads.* Every participant noted that their initial approach on the accessibility forum was to use keywords to search within a thread because manually browsing for information proved to be demanding. Nevertheless, they discovered that the search process was just as burdensome, due to the reasons listed next.

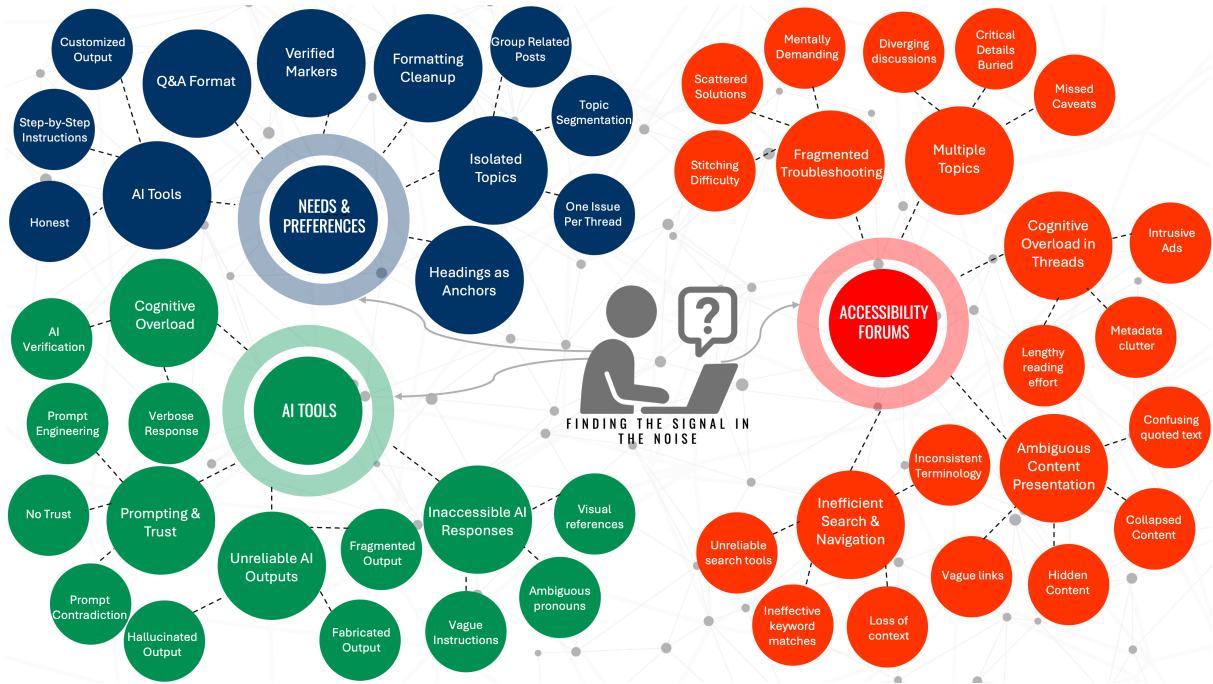

Figure 2: Key themes and sub-themes.

**Broken and unreliable built-in search functions.** Eight out of fourteen participants reported that the search functionality on the *NVDA* and *JFW* forums was unreliable or did not function as intended. As illustrated in Figure 3, the forums lacked thread-level search, offering only site-wide search functionality. Although a search feature existed on individual thread pages, the keyword search often yielded a set of entirely unrelated threads rather than highlighting the relevant terms within the current thread. As a result, participants often became disoriented or confused, requiring considerable time to navigate back to their original workflow. *P1*, for instance, described their experience:

"*Um....you know, search feature is kinda useless there [JWF forums]. I type something in the search box...thinking it will just show me where it is [keywords] in this thread, but instead it takes me somewhere else with a whole list of different threads [search results]. Then I have no idea how to get back to where I was reading.*" – P1

**Burden of linear search keyword navigation.** Consequently, participants often relied on the browser's native search functionality (*CTRL + F*) to locate information within the thread. Yet, this strategy was burdensome because forum conversations were highly intertwined: the same word often appeared repeatedly in quoted passages or tangential replies. For sighted users, irrelevant hits could be quickly dismissed with a glance at the surrounding text. In contrast, *P4* and many other participants (n = 11) noted that they had to listen to each search instance in sequence to determine its relevance, forcing them to wade through numerous false positives without any efficient way to filter:

"*I do Control + F....the search works, but it's not smart. It finds the word everywhere...half the results are just inside someone quoting another post, and I waste so much time checking matches that don't help. It's exhausting.*" – P4

**Guessing the "right" search keyword.** Additionally, forum discussions often contain inconsistent terminology, as users frequently employed different wordings or variations to convey the same information. Blind users, already slowed by linear searching, were further disadvantaged if they failed to anticipate all possible keyword variations. In particular, a few participants (n = 4) reported that it was sometimes challenging to guess the keywords or that they did not know which keywords to search for. As P3 explained: "*I never know what word to type. Sometimes it's 'setup,' sometimes it's 'install,' or they'll call it 'config.' If I don't guess the exact word they used, the search shows nothing useful.*"

**Losing context after a keyword match.** Even when participants successfully located a keyword within a thread, they often struggled to maintain a sense of context. Since they couldn't visually glance above and below a highlighted term in the discussion thread to quickly judge relevance, they reported that search results dropped them directly onto a word in isolation. Subsequently, they had to review the text line-by-line to determine who was posting, their position in the thread, and whether the mention related to the original problem or an off-topic discussion. This constant need to reorient created a heavy interaction overhead, often forcing users to restart listening from the top of the thread. As *P6* described, "*When I jump to a search result, I don't know who wrote it or what part of the thread I'm in. I end up going back to the top just to figure out the context again.*"

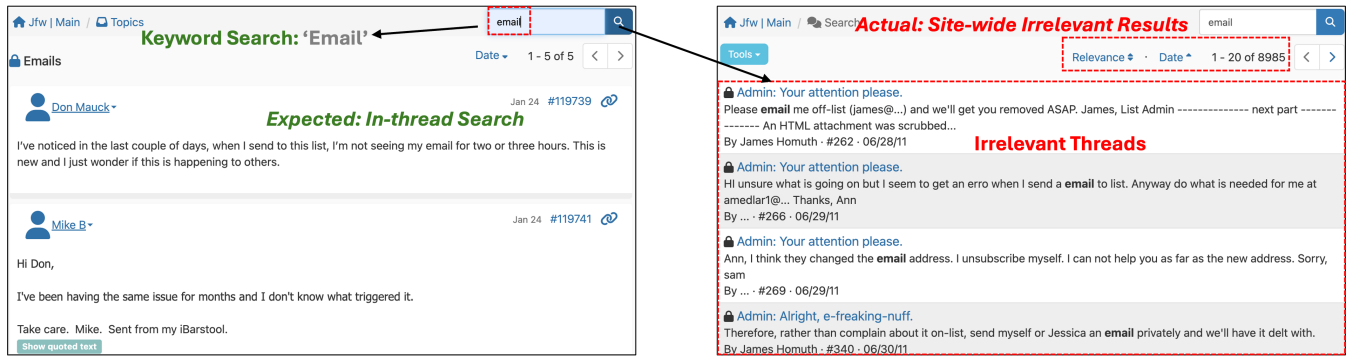

Figure 3: Broken search experience on JFW forum. A keyword search for "email" (left) was expected to highlight results within the active thread, but instead triggered a site-wide search (right), returning thousands of irrelevant posts.

*4.1.2 Unclear Links, Hidden Content, and Quoted Text.* Unclear links, hidden posts, and cluttered quotes emerged as persistent barriers in forums. Instead of aiding navigation, these features often obscured context and made information-seeking laborious for blind users.

**Difficulty with quoted text.** One recurring obstacle in forum use was the presence of quoted text, which further complicated comprehension and navigation. Nearly all participants (n = 12) noted that *NVDA* and *JWF* forums often included passages from earlier posts (see Figure 4A). Some participants found the "Show Quoted Text" and "Hide Quoted Text" buttons occasionally helpful in distinguishing original content from quotes. Nevertheless, many posts contained verbatim copies of previous posts merged into replies. Screen readers read these literally, without clear separation from new content, making it difficult to discern quoted from new material. Participants often had to re-read duplicates, misinterpret conversation flow, or miss new insights. As *P1* remarked, "*I usually skip around, but with quotes, it gets messy. I can't always tell what's new and what's copied, so I end up reading the same thing repeatedly.*"

Additionally, nine participants noted that long, email-like quoted chains created walls of redundant content rather than useful context. As *P7* explained, "*It's useful if it's just the parent post content, but sometimes the whole thread is quoted repeatedly...it becomes noise...like a long email thread.*"

**Hidden content and missed posts.** Beyond issues with quoted text, participants indicated that forums like *Reddit* frequently collapse long discussions or reviews behind expandable sections, further complicating information access. Those who used *Reddit* reported that "View More Comments" button was difficult to access, as screen readers failed to announce it because it was not properly labeled. As a result, participants often missed crucial content or remained unaware that additional posts were available. As *P13* explained:

"*So I was looking up reviews for a new screen reader add-on. At first, I only saw the positive reviews. I didn't realize the negative ones were hidden under a 'view more comments' button, because my screen reader didn't pick it up. So I completely missed them.*" – *P13*

**Ambiguous and unclear link labels.** Every participant reported routinely using screen reader quick navigation keys (e.g., *Insert+F7* to view all page links) to browse and access forum content. While they often listened linearly from top to bottom, skimming was guided primarily by link text. Descriptive links like "Download User Manual (PDF)" were considered useful, but participants often found link labels unclear or confusing.

Seven participants noted significant difficulties accessing links within forum posts, even when listening to the surrounding text. Vague link labels, such as 'here' or 'more', were particularly confusing and frustrating. Participants explained that screen readers announced these literally, without context, leaving them uncertain about what the link referred to or where it led. Consequently, they usually hesitated to follow links for fear of losing their place or opening irrelevant content. As *P9* put it:

"*When the link just says 'here' or 'more,' my screen reader reads out 'link here, link more.' That doesn't make any sense to me. Then I have to stop, go back, and listen to the whole thing again just to figure it out. It gets really tiring.*" – *P9*

**Challenges with embedded links.** Links embedded within sentences were also identified as challenging to access for some participants (n = 6). Screen readers often announced embedded links as simply 'link', without describing the destination. Several participants explained that they frequently did not realize a link was present or what it pointed to. Accordingly, they were often reluctant to click on unfamiliar links, which caused them to miss opportunities to access supporting materials. *P7* recounted, "*I was reading this post about fixing a screen reader thing, and then in the middle it just said 'link.' I'm like, okay, link to what? I had no idea it was the documentation. I only found out later when someone told me. I skipped right over it 'cause I didn't know.*"

*4.1.3 Comprehension Challenges in Mixed-Topic Threads.* Participants described perusing forum threads as an exhausting marathon rather than a simple search for answers. Clutter from metadata, ads, and redundant content turned comprehension into a time-consuming and frustrating task. As a result, staying oriented in the main discussion was difficult, and participants frequently lost track of where critical instructions appeared.

**Distractive metadata.** All participants emphasized that navigating lengthy forum threads and grasping the main discussion was laborious and mentally demanding. They noted that listening sequentially to a single thread with 20 posts could take 25 to 30 minutes. Metadata such as timestamps, usernames, and role markers on each post interrupted the flow and made it even more difficult to focus on the discussion. Participants highlighted that screen readers add

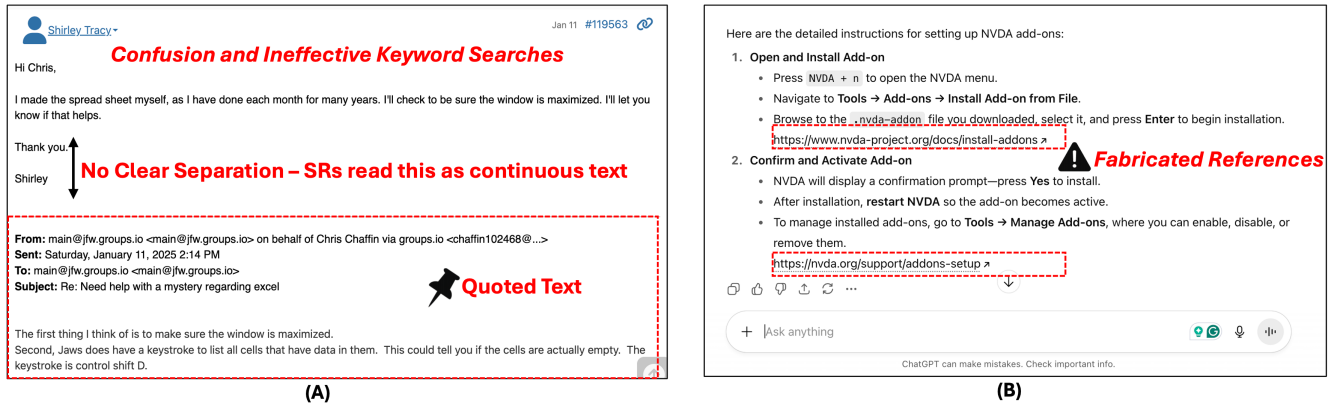

Figure 4: Two barriers to information access. (A, left) Forum replies often embed quoted text without clear separation, causing screen readers to read it as continuous content, leading to confusion and ineffective keyword searches. (B, right) AI-generated answers sometimes fabricate references, which appear credible but are false, forcing blind users into time-consuming verification cycles.

structural and formatting announcements like "heading level 2," "link," or "button," creating additional clutter. This made it easy to "lose track of the actual conversation." *P1* said, "*Too much mess, too much mess...like it has too much information. It's not clean.*" When asked, *P1* elaborated:

"*Yeah, it means...it may have, like, a five-star reply, and then something else that says 'posted two years ago,' and then it says whatever the topic is....username...like this post was posted by Bill. And if the topic is really long, it just goes 'read blah blah blah.' I just wish there was a better way to organize the information to get to the answer faster, especially in accessibility forums like NVDA and JAWS.*" – *P1*

**Disruption from advertisements.** In addition to metadata challenges, nine participants described inline advertisements, pop-ups, and promotional banners in threads as highly disruptive. Since they read posts sequentially, screen readers announced ads with the same prominence as content, forcing extra effort to skip irrelevant material, often multiple times per session. *P2* remarked, "*I'm trying to follow the thread, and suddenly it's reading an ad about something totally random. It's annoying, because I lose the point where I was.*" Similarly, *P13* explained, "*I see the ads all the time. Ads are so annoying! And sometimes, like, there are pop-up video ads that you have to close. That's really frustrating.*"

**Ad Blockers as a complex workaround.** When prompted about ad blockers, participants noted that configuring them was onerous, with different browsers requiring varying procedures to enable. As a result, most did not prefer to use ad blockers, and a few never used them (This is in agreement with findings of Kodandaram et al. [65]). As *P1* remarked:

"*You still have to be a little geeky to use it, because you've got to remember to go into cookies, and then go into this and go into that. And depending on [which] browser, you have to remember different ways to navigate. There's no...it just doesn't work the same everywhere. I feel like there should be a uniform way across browsers...like one shortcut key to disable that same annoying stuff.*" – *P1*

A few participants (n=3) who had experimented with ad blockers further explained that the tools themselves could introduce new frustrations. Occasionally, the blocker would interrupt their workflow by automatically opening a new tab announcing, "We have successfully blocked *X* ads, please rate us, donate, or upgrade." These tabs often contained lengthy forms requesting card details and personal information. For blind users, these unsolicited interruptions were disruptive and disorienting, often breaking focus while reading or navigating threads. As a result, participants reported uninstalling the blockers altogether, noting that once removed, it was not straightforward to reinstall or reconfigure them in practice.

Participants also highlighted that some websites(e.g., *YouTube*) explicitly disallow access when an ad blocker is enabled, forcing them to disable or remove the tool entirely. This created additional barriers, since toggling blockers on and off across sites required extra navigation steps that compounded the cognitive load.

**Multiple Topics Obscure Caveats and Critical Information.** Participants noted that threads often drifted into overlapping discussions, starting with a focused accessibility issue but branching into tangents and sub-questions. This drift made it difficult to pinpoint and extract the specific information they were seeking.

Several participants described how a single question frequently generated multiple solutions, correction chains, and tangential follow-ups. Because screen readers enforce linear listening, participants had to move through every reply to determine which solution applied to their software version or setup. As *P8* recounted their experience:

"*So I was following this thread about a JAWS script error, right? The first reply looked like the solution, but then the next person said, Oh no, that only works on older versions. And then another comment was like, if you're on Windows* 11*, you gotta do this instead. So I couldn't just stop at the first answer — I had to go through all the follow-ups [to figure out what applied to me].*" – *P8*

Participants were also frustrated by off-topic comments that broke their concentration and sometimes led them to abandon a thread entirely. Although accessibility forums have moderators responsible for keeping discussions on-topic, digressions remained common and added noise that made it harder to stay oriented while navigating threads with a screen reader.

Even when listening through threads carefully, participants often missed important caveats that were embedded mid-reply. Multi-topic posts, mixed with quotes, side comments, and partial solutions, obscured corrections or version-specific guidance, making them difficult to notice even when participants believed they had listened to all the follow-ups (see Figure 1). As *P3* reflected: *"Sometimes the important detail is just kind of slipped into the middle of a reply, like, 'Oh yeah, but only if you're on this version.' It doesn't sound like the main point, so unless I'm really paying attention, I don't even realize it's a warning."*

*4.1.4 Mentally Stitching Fragmented Information Makes Troubleshooting Exhausting.* Forums rarely present complete solutions in one place. Instead, blind users must mentally stitch together scattered posts and maintain their own notes, turning troubleshooting into an exhausting, stop-and-start process.

**Fragmented solutions across posts.** Participants consistently regarded accessibility forums as their primary resource for troubleshooting, valuing solutions from individuals with firsthand experience. However, the disjointed and dispersed nature of discussions made it difficult to navigate threads and extract relevant information. As *P1* stated:

"*I prefer it [forum] because there are human beings, and it just seems more human to me, right? Right? It is always best to get solutions from real people who have actually experienced the issues. However, the interface is very messy.*" – *P1*

Several participants (n = 10) noted that solutions were often fragmented across multiple posts, follow-ups, replies, and sub-threads. No single post offered a complete solution, requiring them to mentally assimilate information before applying it in practice.

**External note-taking as a workaround.** Participants likened troubleshooting accessibility issues to 'do-it-yourself (DIY)' tasks, where prior steps must be completed before proceeding. Consequently, they preferred consolidated 'step-by-step instructions' presented in one place. They emphasized that solutions were most effective when clearly articulated, one step at a time, allowing them to follow along without inferring missing information. As *P2* explained:

*"Step-by-step instructions are much better than having a summary. Yes, because it's easy to apply in any practice. Simple instructions, such as 'Do this, then do that,' are easier to follow than verbose explanations."* – *P2*

Similarly, *P14* stated, *"I'm a step-by-step person. Even the official screen reader documentation utilizes step-by-step tutorials to teach users features and customization options. That's how I learn best."*

Participants (n = 8) maintained external notes to consolidate instructions in one place, simplifying troubleshooting. Some copied steps into editing tools like *Notepad* and *Word*, while others used *Braille* notetakers. Nevertheless, note-taking with screen readers remained cognitively demanding and time-consuming, requiring repeatedly switching tabs, cross-referencing posts, navigating back, and re-reading content. *P2* further illustrated this challenge, stating:

*"When I'm troubleshooting, I copy the steps into Notepad so I can keep them in one place. But it's messy. I'm [switching tabs] back and forth — copy, paste, then try to find where I was. Sometimes I lose the thread completely and have to [scroll back] and listen again. It gets tiring real fast."* – *P6*

*P14*, who is a *Braille* user, described, *"I'll try to write things down on my Braille notetaker, but it's slow. I'm listening, then [pausing the screen reader], typing one step, then going back to the post. By the time I finish, I've forgotten what came next, so I have to [re-read] the whole instruction again. It feels endless"* – *P14*

## 4.2 RQ2: Experiences and Challenges with GenAI Tools

Blind users stated that they extensively rely on GenAI tools to manage everyday challenges, including learning and troubleshooting computer-interaction issues. Nonetheless, we observed that these tools introduce a new set of challenges that differ from those blind users experience in accessibility forums, as listed next.

*4.2.1 Inadequate Responses.* AI outputs often failed to provide usable guidance for blind users. Participants reported responses that relied on visual cues, vague pronouns, or underspecified steps, leaving them stranded mid-task and forcing repeated cycles of trial and error, as covered by the following uncovered themes.

**Visual references that blind users cannot access.** A persistent hurdle participants described when using GenAI tools like *ChatGPT* and *Gemini* was that responses were often inaccessible. When addressing accessibility issues, they uploaded screenshots with queries instead of typing questions. As shown in Figure 5, despite explicitly mentioning they were blind, AI outputs frequently relied on visual context, such as "click the button on the top right," "as shown in the screenshot," or "click the green button," which participants could not follow. These descriptions left them uncertain about the referenced element, preventing the successful execution of instructions. *P12* described, *"Even when I tell it I'm blind, sometimes it says, 'look at the screenshot' or 'press the random green button on the screen,' but I don't see any of that. My screen reader just reads a bunch of text. So I'm left guessing which control it actually means."*

**Ambiguous pronouns and context references.** In addition, AI outputs frequently relied on ambiguous pronouns and contextual references, such as "do this," "navigate there," and "select and tick that option." Without any clear anchoring to specific labels, names, or shortcut keys, participants struggled to map the output onto their screen reader workflows. Unlike forums, where human contributors often specify exact keystrokes or menu paths, AI responses risked leaving blind users stranded mid-task. As *P4* explained:

*"It'll tell me 'just click here' or 'navigate to that,' and I'm sitting there thinking, here where? that what? It's not obvious without sight. I need the exact name of the button or the shortcut, otherwise I'm stuck."* – *P4*

**Vague and underspecified instructions.** Participants also highlighted that AI responses were often vague or underspecified, providing superficial advice without actionable details. Phrases like "configure your settings," "navigate to the options menu," or "adjust the preferences accordingly" left them uncertain about the exact menu, tab, or control referenced. While sighted users could infer meaning from the interface layout, blind users needed explicit, step-by-step directions. Several participants explained that such vagueness forced repeated trial-and-error or required re-prompting the AI with increasingly specific questions. As *P7* reflected, *"It'll say something like 'go into settings and change it,' but which setting? Where exactly? I end up asking again and again just to get the exact*

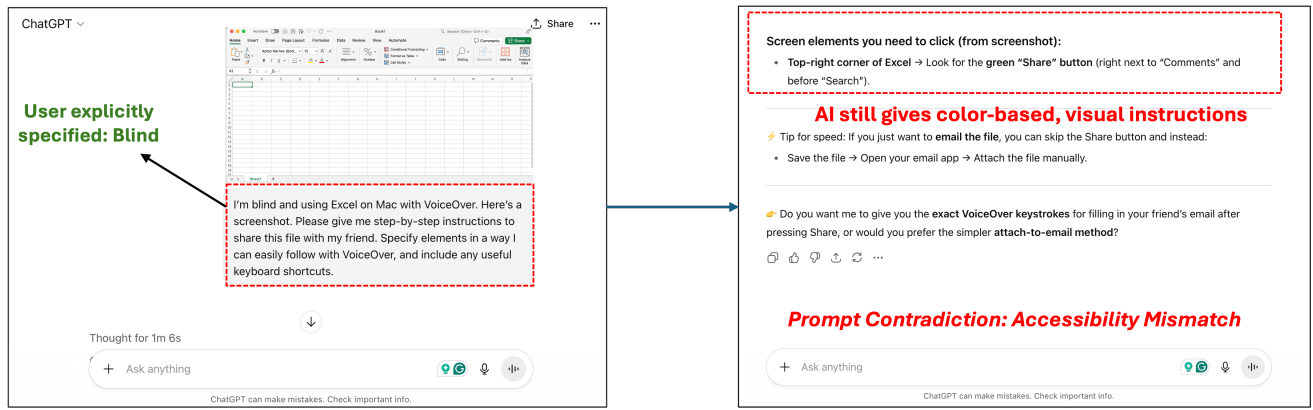

Figure 5: Despite the user explicitly stating they were blind and asking for screen-reader-friendly instructions to share a document in *Excel*, *ChatGPT* responded with inaccessible visual guidance like "click the green button."

steps. It takes longer than if someone had just told me the keys in the first place."

4.2.2 *Unreliable and Misleading AI Responses.* Participants appreciated that GenAI tools like *ChatGPT* generate quick and direct feedback but consistently noted their unreliable responses, as captured by the following themes.

**Hallucinated guidance and fabricated credibility.** Participants reported that AI sometimes produced incorrect troubleshooting guidance, including non-existent menu paths, buttons, settings, and even fabricated citations or tutorial links (Figure 4B). Although sighted users can often quickly dismiss such errors when the referenced UI element is visibly absent, blind participants described the need to systematically traverse the interface with a screen reader to determine whether the instruction was wrong, which added time and frustration. Fabricated references were similarly costly. Confident formatting (e.g., plausible guide titles, version numbers, and URLs) created an air of legitimacy, prompting participants to open links, cross-check sources, and retrace steps before concluding the references were inaccurate or fictitious. While prior work has documented hallucinations in everyday information-seeking tasks [107], our findings reveal that blind users face a disproportionate and often uniquely challenging burden in verifying information.

**Fragmented instructions.** Even when the content was not false, participants explained how AI often produced incomplete or scattered solutions, requiring users to gather partial solutions across multiple responses. Participants noted that an initial answer might offer a starting point, but essential details would only appear after several rounds of re-prompting. This forced blind users into a process of stitching together scattered pieces of information, echoing the burdens they faced in accessibility forums, but without the benefit of clarifications from other human contributors. *P2* described, "*I'll get one piece of the fix in one answer and the rest only if I ask again. It's like trying to build the instructions myself.*"

**Context Leakage Across Chats.** Participants (n = 6) observed that GenAI tools, particularly *ChatGPT*, sometimes appeared to reuse context from earlier conversations, producing answers that blended instructions for the current task with commands or examples from previous, unrelated tasks. Unlike sighted users who might quickly skim and notice such mismatches, blind participants invested substantial effort in listening to and verifying the steps before realizing that parts of the response referred to a different application or problem entirely.

*P3* shared a concrete example: "*I was asking about JAWS, but the instructions had VoiceOver steps in them. At first, I thought maybe I was missing something, so I went through the menus again and again. Then I realized it was pulling in my older chat where I'd been asking about my iPhone. It just blended the two… that confused me completely.*" –P3

Although AI models such as *ChatGPT* are not expected to retain prior sessions, prior work has documented occasional "context leakage," where earlier conversations influence later responses [13, 40, 87]. This blending was especially disruptive in accessibility troubleshooting, where screen reader workflows vary across platforms. When advice for one tool (e.g., *JAWS* on *Windows*) was mixed with instructions for another (e.g., *VoiceOver* on *iOS*), participants could not simply skip irrelevant steps; instead, they had to test, backtrack, and verify each instruction, prolonging troubleshooting and thereby eroding confidence in GenAI tools.

4.2.3 *Context Loss and the Burden of Repetition.* Since GenAI tools generally do not retain information context across chat sessions, numerous participants ($n = 7$) described frustration that they had to repeatedly restate their setup, including screen reader type, version, operating system, and browser, before asking their actual question. This repetitive preamble was exhausting and inefficient. As *P3* explained, "*Every single time, I have to say I'm using JAWS 2024 on Windows with Chrome. By the time I finish telling it [ChatGPT] about my setup, I'm already tired, and I haven't even asked my real question yet.*"

Although GenAI tools like *ChatGPT* now offer features such as Memory, Custom GPTs, and Plugins, which enable users to retain context and receive tailored responses via custom prompts without losing information from previous interactions [86], participants noted that accessing and configuring these features was often challenging without technical knowledge. They emphasized that even a small accidental omission of details in the prompt created a lot of problems. Missing context led to incorrect or generic AI instructions, such as *macOS* shortcuts for *Windows*. As *P5* noted, "*If I forget* to say Windows, it happily gives me Mac instructions. If I don't tell it everything, I just get a generic answer that's no use to me."

This left participants in a double bind: either spend time repeating setup details or risk irrelevant advice. Participants stressed that personalization should be a baseline feature; without it, AI support remained unreliable, increasing verification burdens and eroding trust. As *P7* reflected, "*Maybe it can remember things, but I don't have that version. For me, every new session, I start from scratch...my setup, my screen reader, everything.*"

#### 4.2.4 Prompting, Usability, and Over-Generalization Risks.
Blind participants found that working with AI often felt like a gamble. From guessing the "right" prompt to enduring contradictions, errors, and risky overgeneralizations, trust in these tools quickly eroded.

**Struggling to find the "right prompt".** Several participants (n=11) explained that, similarly to guessing words in forums, they often struggled to know the "right way" to ask AI a question. Slight variations in wording could produce vastly different outputs. Correspondingly, participants would re-prompt and refine phrases to the AI with clarifications until it produced something worthwhile. *P1*, who self-proclaimed to be a technical geek, commented:

"*I have learned. It really depends on how you speak to it. Yeah, to get the information you want from it. It's very good at remembering, but if you don't ask it properly, it will not give you the right information. Something to do with prompt engineering.*" – *P1*

This challenge of formulating effective prompts has also been observed among blind users in the context of everyday practices [6], suggesting that difficulties with prompt articulation are not unique to troubleshooting scenarios but part of a broader accessibility concern in general.

**Prompt contradictions and missed step-by-step instructions.** Numerous participants (n = 9) limited their use of GenAI tools due to negative prior experiences with them. Some (n = 5) reported encountering prompt contradiction [92], a form of AI hallucination in which the system disregards explicit instructions in the prompt. They noted that despite prompting the AI to return step-by-step instructions, the AI frequently produced paragraph-style responses, amalgamating multiple actions into a single sentence. For blind users, omitting even one action meant losing track of the workflow. As *P5* described, "*It'll say, 'open settings, click accessibility, then change the option,' all in one go. By the time I finish the first part, I've already forgotten the rest, so I have to backtrack and repeat.*"

**Frustration with incorrect AI responses.** A few participants (*n* = 4) recounted that their early experiences with GenAI tools were discouraging, as the systems confidently produced instructions that were simply wrong for computer interaction–related tasks. Unlike vague or underspecified answers, these responses were concrete yet inaccurate, creating wasted effort and frustration. For example, *P1* noted:

"*Sometimes I ask about Braille shortcuts in Word or Excel. It'll give me something that looks detailed, but I know for a fact the shortcut doesn't exist. That's not just unhelpful, it's misleading.*" – *P1*

Others described following AI-generated suggestions only to discover they failed in practice. *P6* explained, "*I tried the steps it gave me for fixing a JAWS setting, but nothing happened. I repeated it two or three times thinking maybe I missed something, but no… it was just wrong.*" Such repeated inaccuracies discouraged participants from turning to AI for technical support, especially when checking and re-checking every step with a screen reader already demanded sustained focus and effort.

**Over-Generalized and risky advice.** Additionally, participants emphasized that GenAI tools often overgeneralize; when uncertain, the system still produced answers, defaulting to generic advice like "restart," "reset to defaults," or "update drivers." These responses were frequently outdated, inconsistent across versions, or contradictory across queries. Such advice was time-consuming to verify and risky to attempt, given the effort needed to recover accessibility settings. As *P8* reflected: "*If it doesn't know, it should just say so. Don't give me a confident guess like 'just reset it.' I won't do that....it could wipe out the settings I rely on, and getting back to my setup is a whole project.*"

#### 4.2.5 Cognitive Overload in Consumption and Verification.
Another set of challenges centered on how participants interacted with GenAI tools themselves. Participants (n = 4) found that AI often produced lengthy summaries, verbose responses, even after prompting to return short, step-by-step instructions. They mentioned that it was challenging to find the critical steps that were buried in paragraphs of background. As a result, they had to listen sequentially, sometimes replaying sections multiple times to extract the relevant actions. As *P5* explained, "*It gives me this long story, and the steps are hidden in the middle. I can't just skim, so I listen, rewind, listen again. It takes forever just to find the one thing I actually need to do.*"

Participants emphasized the burden of verifying AI-generated instructions, describing it as a multi-step, time-intensive process: executing each instruction, listening to screen reader feedback, and often retracing steps to ensure nothing was broken or altered. This verification was especially taxing when AI instructions were incomplete, vague, or contradictory. Even accurate guidance required blind users to re-listen and cross-reference with their screen reader. Tasks that take seconds for sighted users could extend into minutes or longer. As *P7* explained:

"*I can't just, you know, glance and check if it worked. I have to [actually] try the step, then wait for the screen reader to read it back...and sometimes I end up undoing it just to be sure. By the time I figure out if it's right....I've already burned so much energy on it.*" – *P7*

### 4.3 RQ3: Needs and Preferences

We asked participants to share their views on how an ideal forum or AI tool should function in practice and what specific needs it should address, in the context of providing computer-interaction support for screen reader users. Our analysis results in this regard are provided next.

#### 4.3.1 Forums: Headings as Anchors.
All participants emphasized the crucial role of headings as structural elements in forums. Unlike sighted users who can scan visually, blind users rely on screen reader shortcuts. Headings serve as navigational landmarks, enabling efficient jumps between posts, sections, or topics without linearizing the entire thread. Participants strongly favored forums where each post is marked as a heading, citing *AppleVis*, which provides a separate heading for every post, enabling direct navigation. They advocated standardizing this practice across accessibility forums. To support it, they proposed attaching *HTML* labels (e.g.,

`aria-label`) to elements so screen readers can announce new topics or posts. They also suggested sub-headings for quoted text, code snippets, or moderator-flagged solutions to enable precise skimming. Without heading markup, participants had to navigate linearly through every post, making orientation difficult. As *P3* explained: "*When there are no headings, it feels like I'm stuck in a wall of text. I don't know where one post ends and the next begins.*"

### 4.3.2 Forums: Accessible Links as Landmarks.
All participants reported relying on screen reader shortcuts (e.g., pressing 'K' in *NVDA* to jump between links) to efficiently browse forum content, making links critical landmarks. As noted in section 4.1.2, vague link text like "click here," "above link," or "below link" forced backtracking or guessing, adding cognitive load. As *P9* explained: "*If the link just says 'click here,' I have no idea where it goes. I have to open it, then come back if it's wrong. That wastes so much time.*"

Participants preferred self-contained, descriptive links with the resource name and URL in brackets, e.g., "Download *NVDA* 2024 installer (nvaccess.org/download)." They also suggested moderators edit or annotate unclear links to ensure accessibility. Clear link labeling reduces frustration and aligns with best practices already encouraged in accessibility guidelines like WCAG.

### 4.3.3 Forums: Streamlined and Clean Auditory Access.
Participants emphasized a strong preference for streamlined access to forums, focusing only on the core conversation (original posts and replies). They were frustrated by extraneous material such as advertisements, repeated metadata (e.g., posted two years ago, usernames, etc.), and verbose signatures with contact information, slogans, or disclaimers. Unlike sighted users who can visually skim past such clutter, blind users must listen sequentially with screen readers, which magnifies fatigue and slows comprehension. As *P5* put it: "*I just want the actual content, nothing else. No ads, no signatures, no repeats — just the real discussion so I can keep up.*"

Raw, unprocessed text also created unnecessary auditory noise. Inline code snippets or ASCII art, for instance, disrupted flow as screen readers verbalized every symbol and space literally (e.g., "space, space, slash, slash, asterisk"). To address these challenges, participants suggested signature-collapsing features, semantic markup for code (e.g., <code> or <pre> tags that trigger the character model), and toggles to skip verbose blocks unless intentionally expanded. Overall, they wanted a cleaner auditory experience that foregrounds discussion content while minimizing redundant or distracting elements in playback.

### 4.3.4 Forums: Isolated Topics.
A single thread often drifts into multiple discussions. For example, a thread might start with, "How do I get *NVDA* to read *Chrome* menus?" but later a user adds, "I also have issues in *Firefox*," followed by a debate on keyboard layouts. For blind users, these replies appear as an undifferentiated stream, making it difficult to determine which posts pertain to *Chrome* and which to *Firefox*.

All participants emphasized that a discussion thread should ideally focus on a 'single' issue. However, they acknowledged that forums are dynamic, and multiple related issues can naturally emerge, sometimes clarifying a problem. A strict one-problem-per-thread expectation is unrealistic. The real need is isolating problems rather than eliminating them. Participants highlighted the importance of topic segmentation and of explicitly marking or grouping related posts, which they noted would reduce confusion and enable more efficient navigation.

A few participants (n = 5) expressed frustration that topic segmentation has long been discussed, yet mainstream forums have not implemented it. They noted that some third-party Chrome extensions attempt to address this issue, but these are either time-limited or require paid subscriptions, which participants found discouraging. As *P6* remarked:

"*It's very disappointing that nobody has implemented this [topic segmentation] yet, not even in accessibility forums. I have requested this feature many times [to JFW], haven't heard back from them.*" –P6

### 4.3.5 Forums: Q&A Pairs as a Preferred Format.
Several participants (n = 8) expressed a strong preference for forum content structured as Question–Answer (Q&A) pairs. While topic segmentation aids navigation, it still leaves the burden of searching long threads. Q&A pairs would distill the discussion: the key problem framed as a question, followed by ordered solutions and explanations.

Participants envisioned solutions as step-by-step instructions. This structure allows quick identification of actionable guidance without linearizing unrelated digressions. As *P2* explained: "*I don't want to wade through twenty posts to figure out which ones are useful. Just give me the main question and then all the good answers, in steps I can follow.*"

Additionally, Participants emphasized that if a question has no available solution, the thread should explicitly mark it as "no solution." This would save considerable time otherwise spent searching through posts. As *P4* noted: "*I'd rather be told straight away that there's no fix than keep scrolling through posts hoping to find one. Sometimes I waste half an hour trying every suggestion...only to realize at the end that nobody actually had a working answer. Just mark it as 'no solution' so I know not to bother.*"

### 4.3.6 Forums: Verified Markers Reduce Troubleshooting Burden.
Participants explained that troubleshooting in forums carries a disproportionately high burden for blind users. Verifying solutions is time-consuming, as each step must be executed carefully, interpreted via the screen reader, and often undone if it fails. As *P1* noted, "*I can't just glance and see if it worked. By the time I know, I've already spent so much effort.*"

To reduce the burden of "high cost of verification," participants strongly desired markers for verified information. They suggested confirmations from the original poster, endorsements from developers, or community validation where multiple users confirmed success. As *P2* explained, "*If I knew right away which answer worked for others, I wouldn't waste so much time testing things that go nowhere.*"

Several technically experienced participants pointed to StackOverflow's green tick marker and Apple Discussion Forum's top-ranked reply as models that quickly highlight reliable solutions. They noted that accessibility forums, despite higher verification costs, lack such mechanisms. Incorporating verified markers, semantically accessible for screen readers, would significantly reduce effort and allow blind users to focus on the most trusted, actionable responses immediately.

*4.3.7 GenAI Tools: Honest and Reliable Output.* Many participants (n = 9) stressed that when troubleshooting with AI, they wanted the system to be transparent about its limitations. They preferred outputs that explicitly flagged uncertainty and offered cues to judge reliability, rather than authoritative-sounding answers that might be incorrect.

Participants emphasized honesty over fluency. AI should indicate when it is unsure, outdated, or drawing on incomplete knowledge. As *P1* noted: "*I'd rather it say 'I don't know' than give me something that sounds right but isn't. At least then I know not to waste my time.*" Others highlighted the value of contextual qualifiers like version numbers or recency markers. As *P7* commented: "*If it said, 'This is for JAWS 2021,' or 'might not apply to the latest NVDA,' I'd know right away if it's relevant to me.*"

*4.3.8 GenAI Tools: Preference for Structured Step-by-step Instructions.* A majority of participants (n = 10) preferred AI responses in step-by-step, structured formats rather than long, conversational paragraphs. They noted that unstructured text is difficult to process with a screen reader. As *P2* explained, "*When it gives me five things in one sentence, I lose track. Just give me step one, step two, step three...that's what works.*"

Participants highlighted the value of embedded checkpoints signaling success or failure, reducing guesswork. As *P9* commented, "*I want it to tell me, 'If it worked, you should hear this.' That way, I know right away instead of guessing.*"

They also preferred responses structured with lists, headings, or navigation markers recognizable by screen readers, enabling efficient skimming and jumping between parts. As *P5* explained, "*Lists are really helpful because I can jump between them with my screen reader. A giant block of text is almost impossible to skim.*"

*4.3.9 GenAI Tools: Customization for Individual Setups.* Over half of the participants (n = 8) emphasized that AI should adapt to their specific screen reader and browser combinations, as generic instructions often misaligned with their setup, forcing extra troubleshooting. As *P4* explained, "*It tells me to click or use a shortcut that doesn't even exist in my screen reader [JAWS]. Then I have to figure out if it's for JAWS or something else.*"

Participants preferred customizable profiles specifying their environment (e.g., "JAWS with Chrome" or "NVDA with Firefox") so responses could be tailored. As *P6* noted, "*If I could just tell it once that I'm using JAWS with Chrome, it should remember and give me answers for that, not VoiceOver or something else.*"

Environment-specific guidance was seen as critical to reducing wasted effort. As *P12* explained, "*One wrong key command costs me time. If it gave the right shortcut for my screen reader, I'd trust it a lot more.*" These accounts underscore the importance of context-aware assistance over generic troubleshooting.

## 5  Discussion and Future Research Directions

Our research examined how blind users interact with accessibility forums and GenAI tools to learn, configure, and troubleshoot assistive technologies. While prior studies highlighted the potential of these platforms for information retrieval, our findings reveal persistent challenges, including the cognitive burden of piecing together fragmented forum content and navigating inconsistent AI outputs such as fabricated information, which are amplified when using screen readers. Next, we situate these findings within prior work, outline design implications for future systems, and highlight our study limitations and directions for further research.

### 5.1  Contextualizing Our Findings vis-à-vis Prior Related Studies

Below, we situate our findings in the context of prior accessibility and GenAI research.

*5.1.1 Broadening Understanding of Forum-Based Strategies Among Blind Users.* Our findings significantly deepen current knowledge of how blind users address computer-interaction needs. Whereas existing work has primarily examined how blind screen-reader users formulate help-seeking queries on accessibility forums, detailing their goals, attempted steps, and system state (*e.g., versions and configurations*) to obtain tailored assistance (Section 4.2 of Saha et al. [99]; Section 4 of Johnson et al. [60]), our study goes further by revealing how they consume and operationalize community responses to troubleshoot their issues. Participants described troubleshooting as a deliberate "do-it-yourself" process, assembling version-specific instructions from multiple posts, tracking corrections, and attaching confirmations to each step. Intertwined conversations and fragmented information often made it difficult to stitch together the relevant steps. To make responses operationalizable, participants adopted various workarounds, such as keeping external notes, copying steps into editors like *Notepad* or *Word*, or using *Braille notetakers*. These strategies highlight the substantial, often invisible effort required to transform forum replies into operationalizable steps for resolving computer-interaction problems.

*5.1.2 Exploring Navigation to Operationalization.* While existing work [11, 106, 126] examined the navigational challenges faced by blind screen reader users, such as navigating between posts and locating solution providing posts within a thread (Section 3.3 of Sunkara et al. [106]; Section 4.6 of Aiyer et al. [11]), our study goes beyond navigation by advancing knowledge of how blind screen reader users extract relevant pieces of information across posts within a thread and stitch them together to form a coherent sequence of steps when troubleshooting computer-interaction issues. Our findings show that, beyond navigation difficulty, inaccessible presentation, including long, intertwined, mixed-topic threads, redundancy, and missing context, exacerbates the challenge of discovering and extracting critical steps, even when correct answers exist in the thread.

*5.1.3 Re-examining the Usability of Accessibility Forums.* In prior work, Venkatraman et al. [111] examined comprehension during screen-reader listening and found that, compared to general forums, accessibility forum posts have higher lexical density and more descriptive action verbs, arguing that this explicit instruction-like language can make posts easier to follow with screen readers (Section 4 of Venkatraman et al. [111]). Our findings showed a more mixed picture. While participants sometimes appreciated concise, action-oriented phrasing, they also reported that dense wording combined with structural issues in thread navigation often hindered comprehension. Long conversations with repeated quotations obscured new information, and metadata such as timestamps, usernames, and role labels punctuated posts in ways that disrupted listening flow.

Prior work also reported that ad blockers often did not function well with screen readers, and that blind users were sometimes unaware of how to install or configure them (Section 3.3 of Kodandaram et [65]). Our participants confirmed these challenges and explained that many avoided ad blockers because configuration was burdensome, procedures varied across browsers, and blocker notifications frequently interrupted workflow. Participants described losing their place when ads appeared mid-thread and sometimes uninstalled blockers after repeated disruptions. Together, these findings show that, although accessibility forums are crucial resources for blind users, their usability is constrained not only by navigation challenges but also by fragmented threads, frequent metadata interruptions, and ad-related disruptions.

*5.1.4 Advancing the Understanding of Blind Users' Experiences with GenAI for Computer-Interaction Support.* Our findings broaden the understanding of blind screen-reader users' experiences with GenAI, expanding the scope beyond everyday applications to include support for computer-interaction tasks. Prior work on GenAI with blind users primarily emphasized day-to-day applications such as visual assistance and content creation, highlighting coarse-grained hallucinations and the verification burden, including fabricated but plausible details, outdated information, and substantial cross-checking with other tools or people (Section 4 of Adnin et al. [6]; Section 4.4 of Tang et al. [107]).

Beyond these everyday scenarios, our study examined computer-interaction troubleshooting and uncovered concrete, task-specific hallucination patterns. Participants encountered prompt contradictions (a hallucination type; see Section 4 of Perković et al. [92]) in which GenAI ignored explicit requests for step-by-step output and returned unstructured prose, as well as sentence-level contradictions with fabricated references and non-existent buttons or menus, increasing verification effort. These responses compelled users to verify information across forums, official documentation, and other GenAI tools. Participants also faced underspecified guidance, including referential ambiguity (e.g., "do this"), visual-only cues that assumed sighted navigation, and missing preconditions such as versions and required states, undermining reliable task execution.

Long, unstructured GenAI responses compounded these difficulties. Unlike sighted users who could skim, screen-reader users listened sequentially and often replayed sections to extract operationalizable steps, which participants described as cognitively taxing and inefficient. They preferred concise, procedural, keyboard-only guidance that specified environment details (e.g., screen-reader version) and included intermediate confirmations before proceeding. These preferences aligned with prior reports of verification demands in everyday GenAI use by blind people [107], and our results situated them in technical troubleshooting and system configuration, where sequential consumption magnified verification costs. Taken together, hallucinations, verbosity, and referential ambiguity increased verification effort and underscored the need for structured, version-aware, confirmation-driven GenAI guidance for reliable troubleshooting.

## 5.2 Rethinking Design of Forums and AI for Accessibility

Our findings offer valuable perspectives for designing intelligent intervention tools and strategies that enable blind users to acquire information for addressing their accessibility challenges more efficiently. Such approaches can support a more productive and less mentally taxing experience. We discuss several key insights below.

*5.2.1 Minimizing Metadata Overload.* Participants noted that excessive metadata, such as repeated quotes, usernames, and timestamps, cluttered screen reader output and hindered comprehension. To balance context with accessibility, we suggest lightweight "reference mechanisms," i.e., instead of embedding full quoted passages, replies could include a compact reference (e.g., "In response to: [Post Title/Heading]"), allowing users to jump to the original content only if needed. Metadata such as usernames and timestamps adds overhead when read aloud. For example, screen readers often announce, "This post was created by Bill at 2:00 PM," for every post, which becomes repetitive in long threads. To reduce this burden, platforms could condense metadata into short phrases (e.g., "Bill, 2 PM"), apply progressive disclosure with toggles or shortcuts, or group metadata in a collapsible header. These approaches maintain provenance while minimizing cognitive load for blind users.

*5.2.2 Making Reliability Visible.* We suggest integrating lightweight markers to indicate when information has been verified, similar to Stack Overflow's green tick symbol [3]. Participants noted that such cues would help distinguish tentative suggestions from reliable answers without reading entire threads. Verification could be initiated by thread originators confirming when a solution worked, reinforced through community endorsements (e.g., upvotes), or validated by moderators and domain experts with a "reviewer" status. To provide nuance, markers should also capture contextual caveats (e.g., "works only on JAWS 2024, Windows 11"), clarifying scope and limitations. An AI-assisted feature could further summarize the consensus between replies, revealing which solutions are confirmed, disputed, or untested. Together, these mechanisms would streamline verification and allow blind users to identify operationalizable guidance more efficiently without exhaustive linear perusal of threads.

*5.2.3 From Chaos to Clarity: Forum Designs for Automatic Topic Segmentation.* Although maintaining a single "topic" per thread is impractical, discussions can be categorized and grouped by content. Topic-driven segmentation is a well-established approach, yet accessibility forums have not implemented it to support blind screen reader users.

Given these issues, forum developers should design platforms that scaffold topic segmentation during post creation. Structured templates could prompt users for metadata such as topic labels, short descriptions, post heading, contextual details (e.g., device/OS/software), and post type (question, follow-up, bug report, workaround). Moderators could verify this metadata, serving as anchors for algorithms to cluster related responses, distinguish primary from secondary issues, and highlight key solutions. Automatic segmentation techniques or LLMs could detect subtopics, flag digressions, and suggest splitting or linking discussions, making threads more coherent and navigable for blind users.

---

[3]https://meta.stackoverflow.com/questions/379675/who-marks-green-tick-in-stack-overflow

If collecting metadata directly from users is difficult, platforms could instead post-process posts automatically, restructuring long posts into labeled sections, reformatting replies into collapsible sub-threads, and generating navigable summaries. This balances user effort with automation while improving accessibility and readability overall for screen reader users.

*5.2.4 Blending Human Validation and AI Synthesis.* Our study revealed that blind users strategically combine both forums and GenAI tools to meet their information needs. Forums are valued for human validation, peer support, and the sense of a trusted community, whereas GenAI tools are appreciated for their immediacy and ability to generate tailored responses when prompts are carefully crafted. Yet, to date, no system has sought to combine these two complementary strengths into a unified solution.

Building on this insight, we advocate the development of an intervention tool, for example, a *Chrome Extension* [4], that re-formats forum data into structures that better align with blind users' preferences. Specifically, the tool should restructure threads into coherent question–answer units, synthesize step-by-step instructions, and smartly surface essential caveats drawn from related clarifications and follow-ups. Since caveats are crucial for effective troubleshooting but are often buried deep in long discussions, their integration would reduce errors and wasted effort.

By leveraging advanced techniques such as recursive thread summarization, clustering, and answer ranking [123, 125, 127], the extension could present reliable, community-validated knowledge in a format that combines the trust of forums with the immediacy of AI. In addition to troubleshooting threads, the tool could adapt to other thread types (e.g., announcements), treating the product or version as the "question" and organizing announced features as structured "answers." Importantly, it should also support non-visual skimming by allowing users to jump directly between questions, bypassing irrelevant sections without linear reading.

*5.2.5 **Reviewer–Refiner Loops with Multi-Agent Systems**.* Our findings show that AI often produces unreliable responses for blind users, including hallucinations and fabricated details (e.g., tutorials), requiring manual verification. This issue has been documented in prior studies [6, 34, 55, 107]. One approach to improve reliability is to leverage large language models within a multi-agent framework [36, 109]. We envision an intervention tool that enables multiple agents to collaboratively generate, review, and refine content before presenting it to users. One agent produces an initial response, while others verify factual accuracy, remove vague or visually dependent instructions, and consolidate fragmented replies into coherent explanations. This reviewer–refiner loop ensures blind users receive outputs that are accurate, actionable, and accessible, reducing the burden of manual verification.

Such multi-agent strategies can generalize to diverse accessibility tasks where reliability and clarity are critical. Recent work on agentic frameworks, including AutoGen [116], CAMEL [69], and Reflexion [102], shows that structured agent collaboration improves reasoning, reduces hallucinations, and produces higher-quality task-specific outputs. Building on these techniques, AI–HCI researchers should design tools that leverage multi-agent reviewer–refiner loops to address accessibility challenges. These tools could automatically filter inaccessible content, flag potentially misleading answers, and restructure outputs into non-visual, step-by-step instructions, providing immediate, practical benefits for blind users.

### 5.3 Contextualizing our findings through the Lens of Information Foraging Theory

Blind users' troubleshooting strategies across accessibility forums and GenAI tools can be viewed through the lens of Information Foraging Theory (IFT), which conceptualizes information seeking as an optimization process in which people aim to obtain "maximum information" while expending "minimum effort" in seeking, navigating, and interpreting the information [94]. Key constructs include information scent (*cues that signal potential usefulness*), information patches (*clusters of related information such as threads or GenAI responses*), foraging cost (*effort required to navigate, find, and extract relevant information*), and foraging strategies, the adaptive behaviors people use to minimize cost, such as changing sources, reformulating queries, and selectively scanning cues.

Our findings show that blind users are often forced into high-cost foraging strategies in accessibility forums due to weak scent and poor patch structure. Weak scent arose from inconsistent terminology, vague link labels, ineffective search, and quoted text blended with replies, obscuring where valuable information was likely to appear. Poor patch structure reflected intertwined threads and mixed topics that made it difficult to locate and extract succinct sequences of operationalizable steps. Consequently, participants adopted compensatory strategies, such as linear traversal, repeated reorientation after search hits, maintaining external notes, and manually stitching fragmented solutions from multiple replies. These strategies reflected attempts to cope with high navigation and consumption costs within patches that offered low reward density.

In contrast, GenAI tools initially offered a stronger scent and thus attracted users as a "lower-effort" starting resource. However, hallucinated steps, fabricated information, visual instructions, and contradictory outputs often imposed substantial verification cost, requiring strategies such as step-by-step validation, iterative re-prompting, and cross-checking GenAI outputs with forum content. As a result, users routinely switched between forums and GenAI tools. They quit forums when navigation became overwhelming, and they quit GenAI tools when hallucinations increased verification costs. This pattern aligns with IFT's prediction that users switch between sources when the perceived reward declines, even when the source contains relevant information but the "Effort-to-Value" ratio remains high.

By interpreting blind screen reader users' troubleshooting strategies through the lens of Information Foraging Theory, several design implications emerge for enhancing information scent, restructuring patches, and reducing information foraging costs. Accessibility forums could use clearer headings, descriptive link labels, distinguishable quoted text, and support summaries, topic segmentation, Q&A pairs, and verified-solution markers to reduce manual effort in locating and assembling relevant steps. GenAI tools could reduce verification load by providing structured, assistive-technology-aware instructions and explicitly signaling uncertainty. From an IFT perspective, improving troubleshooting means creating environments where users can rely on low-cost, high-confidence strategies rather than labor-intensive compensatory ones.

---

[4]https://chromewebstore.google.com/category/extensions

### 5.4 Limitations and Future Work

Our study did have a few limitations. Most participants were intermediate or expert screen reader users, with only two beginners, which may underrepresent novices or those with limited technical skills. We focused on accessibility-specific forums, leaving out general-purpose forums (e.g., Quora) where blind users seek broader information. Only English-language forums and GenAI tools were analyzed, so multilingual contexts may pose additional challenges. Finally, our findings reflect the interpretive analysis of interview data; alternative themes or connections might emerge under different analytical perspectives.

These limitations suggest directions for future research. Studies could recruit a wider spectrum of users, including novices, to capture varied expertise levels. Expanding to general-purpose forums and multilingual environments would illuminate additional barriers and patterns. Longitudinal studies could track changes in forum and AI tool usage over time [32]. Prototyping and evaluating accessibility-aware features, such as semantic post grouping or step-by-step AI response modes, would provide empirical evidence for interventions that reduce verification effort and enhance trust.

## 6 Conclusion

This paper examined how blind users engage with accessibility forums and generative GenAI tools to troubleshoot and learn about assistive technologies. Despite being indispensable resources, both systems impose disproportionate burdens. Forums overwhelm users with fragmented discussions, redundant content, and scattered solutions that must be mentally stitched together. GenAI tools promise immediacy but often generate inaccessible, vague, or fabricated responses that require repeated prompting and verification. These shortcomings turn routine support-seeking into a cognitively demanding process.

Our findings contribute an empirically grounded understanding of blind users' needs, including concise step-by-step instructions, explicit caveats, and trustworthy outputs that reduce verification effort. Building on these insights, we outline design opportunities such as topic segmentation and Q&A restructuring in forums, and multi-agent verification and screen reader–optimized outputs in AI systems. Re-imagining forums and GenAI tools with accessibility at their core can transform them into cognitively manageable and empowering resources, enabling blind users to troubleshoot, learn, and participate more fully in digital environments.

## Acknowledgments

We thank the anonymous reviewers for their insightful feedback. This work was supported by the Google Inclusion Research Award, NSF Award: 2153056, NIH Award R01EY035688, and DoD award: HT94252410098.